# Dry dilution refrigerator with $^4$He-1K-loop.

**Abstract**

In this article we summarize experimental work on cryogen-free $^3$He/$^4$He dilution refrigerators which, in addition to the dilution refrigeration circuit, are equipped with a $^4$He-1K-stage. This type of DR becomes worth considering when high cooling capacities are needed at T ~ 1 K to cool cold amplifiers and heat sink cables. In our application, the motivation for the construction of this type of cryostat was to do experiments on superconducting quantum circuits for quantum information technology and quantum simulations. In other work, DRs with 1K-stage were proposed for astro-physical cryostats. For neutron scattering research, a top-loading cryogen-free DR with 1K-stage was built which was equipped with a standard commercial dilution refrigeration insert.

Cooling powers of up to 100 mW have been reached with our 1K-stage, but higher refrigeration powers were achieved with more powerful pulse tube cryocoolers and higher $^4$He circulation rates in the 1K-loop. Several different versions of a 1K-loop have been tested in combination with a dilution refrigeration circuit.

The lowest temperature of our DR was 4.3 mK.

1. **Introduction**

Dilution refrigerators (DR) are the work horses for scientists doing research at milli-kelvin temperatures [1]. We estimate that over 100 new commercial DRs have been manufactured annually worldwide in recent years. Whereas DRs have been pre-cooled by liquid helium cryostats in the past, modern DRs are pre-cooled by pulse tube cryocoolers (PTC). Since our first publication on this subject [2], pulse tube precooled DRs have been commercialized, and today several cryo-engineering firms offer them for sale in different sizes and with useful options for experimentalists. One important feature of cryogen-free DRs is that they offer lots of experimental space in the region of the mixing chamber. The diameter of its mounting plate can be increased up to ~ 50 cm and higher as there are only radiation shields and a vacuum jacket with these cryostats and no helium dewar. This version of DR is easy to operate, cost efficient, and can be completely computer controlled.

Besides the traditional areas of research like materials research, neutron scattering or astro-physics, quantum information technology has been a field of great interest in recent years where many DRs are employed. For this application we have constructed a cryogen-free test cryostat where in addition to the dilution circuit a separate $^4$He loop with a base temperature of about 1 K was installed to increase the cooling power in this temperature range. This loop reaches cooling powers of up to 100 mW in our cryostat which is a factor of 5 to 10 higher than the cooling power of the still of the DR which usually has to be used for cooling purposes at T ~ 1 K. Even higher refrigeration powers have been reported for a 1K-stage which was mounted in a top-loading cryostat for neutron scattering research; here, a cooling power of 230 mW at a temperature of 1.9 K was quoted [3].

Some of the experimental results described in the following have already been presented at various cryo-engineering conferences.

## 2. Proposed concepts for cryogen-free DR with 1K-stage

Cryogen-free DRs with 1K-stage have been proposed before by Hollister and Woodkraft [4], mainly to cool instrumentation for astronomical and particle physics applications. Three different designs were suggested by these authors to cool a condenser where the $^3$He flow of the DR is liquefied and where additional refrigeration power is available to cool amplifiers and heat sink cables.

One design uses a pair of charcoal adsorption pumps which are alternately cooled by the second stage of a PTC to T ~ 4 K. Those pumps are connected to $^4$He evaporators which can alternately be thermally connected via gas gap heat switches to a 1K condenser which liquefies the $^3$He flow of the DR. While the first evaporator expends its liquid helium supply, the second evaporator is regenerated and vice versa. An external pumping system and a gas handling board would be superfluous with this concept. The problem with it would be that dimensions of the charcoal pumps are getting prohibitively big with rising requirements for the refrigeration power of the 1K-stage.

In another scheme, the $^3$He of the DR is condensed with a $^4$He closed-cycle cooler where the back-streaming helium is pre-cooled by the two stages of a PTC and then expanded in a flow restriction. The liquid fraction of the helium flow would be accumulated in a container where it would be available for cooling. In this concept, a counterflow heat exchanger (cf-hx) in the $^4$He-loop to further precool the helium flow before the expansion was not proposed. To circulate the helium, rotary pumps and a simple gas handling system for gas storage and handling are necessary.

The $^4$He-loops described in the following sections are all of this second type.

## 3. Configuration 1: DR and 1K-stage separate

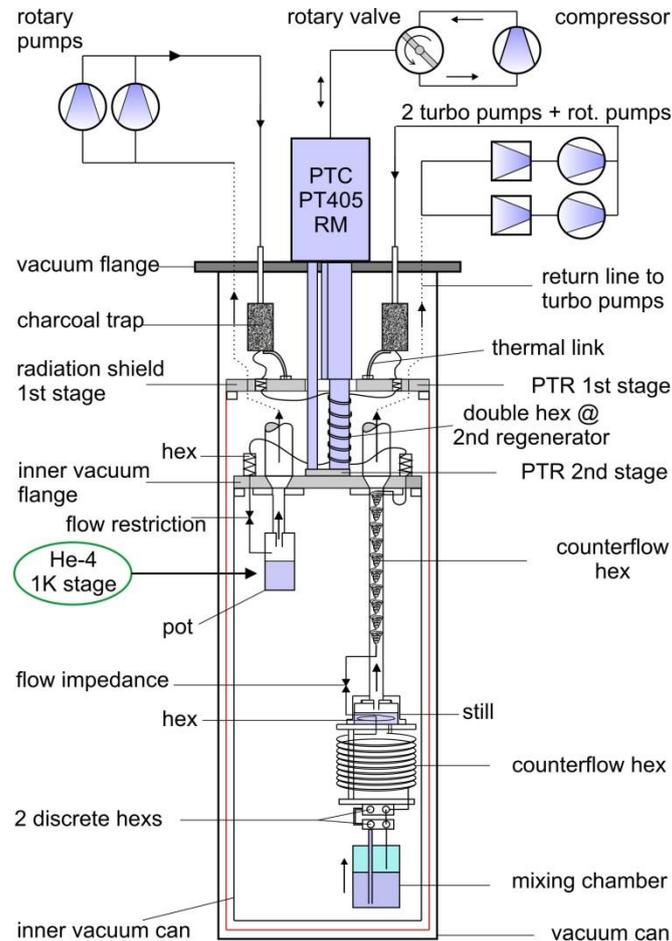

Fig. 1. Cross section of our DR with 1K-stage. In this version the dilution unit and the 1K-loop are separate. The 1K-stage has no counterflow heat exchanger to precool the $^4$He flow; its temperature before expansion is equal to the temperature of the $2^{nd}$ stage of the PTC.

In Fig. 1, a cross section of our cryostat is given; a pulse tube cryocooler (PTC) precools the $^3$He flow of the dilution refrigeration unit and the $^4$He flow of the 1K-loop. We chose a small PTC (0.5 W at 4 K), mainly to keep mechanical and acoustic vibrations low, but also to conserve electrical power [5]. The downside was that cooldowns from room temperature took longer with the small PTC (∼ 20 hours). Temperatures of the PTC were measured with a Cernox resistor [6] and PT100 resistors. In contrast to most commercial dry DRs, this cryostat has an inner vacuum can which is thermally connected to the $2^{nd}$ stage of the PTC via copper braided ropes; thus the cooldown of dilution unit and 1K-stage from room temperature to ∼ 15 K can be done in $H_2$ (or Ne) exchange gas. Pumping of the exchange gas is not necessary after a cooldown; it can remain in the inner vacuum can where it freezes on further cooling. Both vacuum cans are from aluminum, whereas the radiation shield of the 1$^{st}$ stage of the PTC is from copper.

The 1K-loop is depicted on the left side of Fig. 1. A simple gas handling system was set up for helium storage and gas manipulation; to circulate the $^4$He gas, we have tested standard rotary pumps and oil-free scroll pumps with pumping speeds between 30 m$^3$/h and 100 m$^3$/h. A compressor was not required with the $^4$He loop; maximum condensation pressures were always below 0.1 MPa. At the cryostat inlet, the $^4$He of the 1K-loop and the $^3$He of the DR are each purified in charcoal traps which are thermally connected to the 1$^{st}$ stage of the PTC by copper braided ropes. No other cold gas purifiers are used with the gas circuits, so this is a truly cryogen-free DR. Next, the two circuits are cooled in hxs at the 1$^{st}$ stage, at the 2$^{nd}$ regenerator [7,8] and at the 2$^{nd}$ stage of the PTC. The hxs at the 2$^{nd}$ regenerator consist of two capillaries which were soft soldered to the regenerator tube; a photo of the hxs is found in [8].

From here, the $^4$He flow of the 1K-stage can either be expanded in a flow restriction (Fig. 1) or alternatively be run through a counterflow heat exchanger (cf-hx) and a flow restriction (Fig. 4). After the expansion, the liquid fraction of the $^4$He accumulates in a vessel ($V_{vess}$ = 53 cm$^3$), whereas the gas fraction flows back to the pumps. The flow restriction of the 1K-stage which controls the flow rate was made from a piece of capillary (0.1 mm i.d.). In cases where a higher flow resistance was needed, this capillary was pulled through a wire draw plate where the outer and inner diameters of the capillary were reduced.

The refrigeration power $Q_{1K}$ of the 1K-stage is given by the enthalpy balance of the in- and outflowing $^4$He:

$$Q_{1K}/n_4 = H_{4sat}(T_{vess}) - H_4(p_{in4}, T_{in}) \qquad (1)$$

Here $H_{4sat}(T_{vess})$ is the enthalpy of the saturated gas at $T_{vess}$ (vessel temperature) and $H_4(p_{in4},T_{in})$ is the enthalpy of the $^4$He at the inlet of the flow restriction of the 1K-stage (Fig. 1). $p_{in4}$ is the inlet pressure (Fig. 2); the inlet temperature $T_{in}$ agrees with the temperature of the 2$^{nd}$ stage $T_{PT2}$ of the PTC. $n_4$ is the $^4$He flow rate.

In order to calculate the cooling power in Eq. (1) precisely, enthalpy data are best taken from software [9]. For a rough estimate of $Q_{1K}$, $H_4(p_{in4},T_{in})$ can be ignored in Eq. (1); at $T_{in}$ < 3 K and 0.25*10$^5$ Pa < $p_{in4}$ < 1*10$^5$ Pa, the latent heat of vaporization $L_v$ has already been removed from the $^4$He flow and thus $H_4(p_{in4},T_{in})$ is small. Furthermore, at $T_{vess}$ ~ 1 K, $H_{4sat}(T_{vess})$ ~ $L_v(T_{vess})$ ~ 20 J/g. In summary, we have:

$Q_{1K}$ ~ $n_4$ * 20 J/g, where $n_4$ is in g/s and $Q_{1K}$ in W.

This relation is also valid for the case that a cf-hx has been installed in the 1K-loop (Fig. 4).

Superfluid film flow has not been a problem with the 1K-stage, although from time to time minor changes of vessel temperature and $^4$He flow rate were observed. There was no film burner in the present construction to prevent superfluid flow, but it would not have been difficult to build one into the pumping line of the pot. Finally, the inlet pressure $p_{in4}$ and the vessel temperature of the 1K-loop are shown in Fig. 2 as a function of the vessel cooling power $Q_{1K}$. With increasing $Q_{1K}$, the vessel temperature $T_{vess}$ rises and with it the $^4$He vapor pressure and the flow rate, and therefore the inlet pressure also rises.

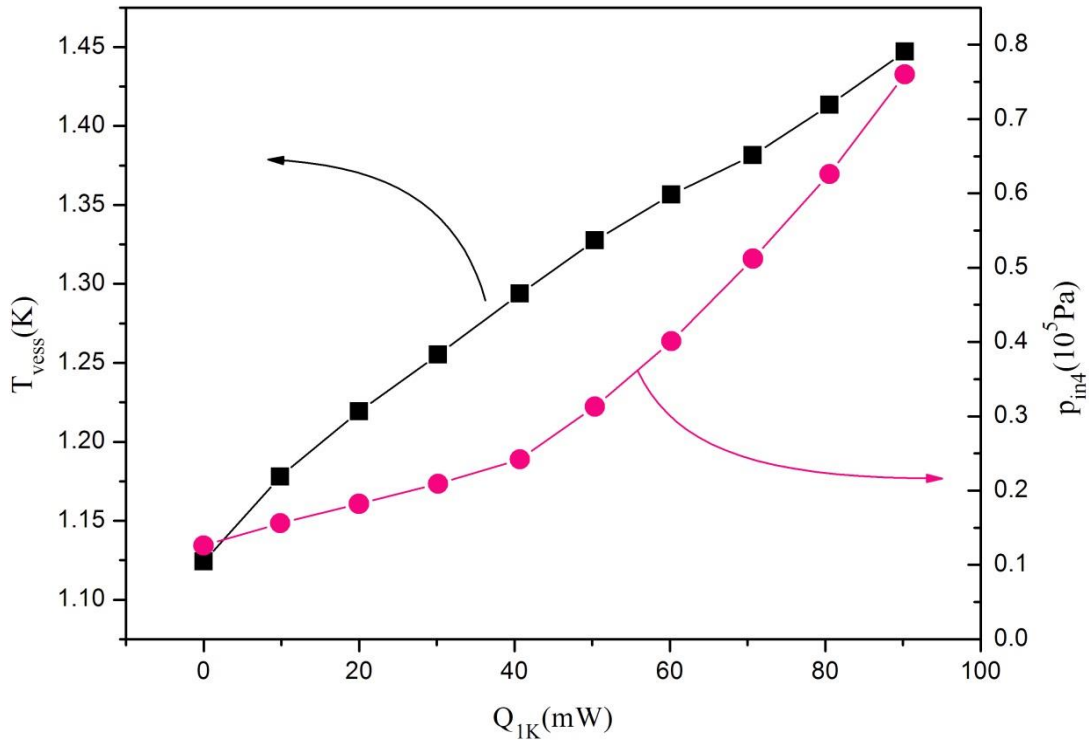

Fig. 2. Vessel temperature and pressure at the inlet of the 1K-stage as a function of the cooling power. For this experiment an 80 m$^3$/h pump was used to circulate the $^4$He.

The dilution unit is depicted on the right side of the cryostat in Fig. 1. Our manually operated gas handling board is not shown in the figure. The dilution loop could be run either alone or together with the 1K-loop. In the DR, the $^3$He was circulated with 2 turbo pumps (Pfeiffer 1600) which were backed either by 2 rotary pumps (Alcatel 2033H) or 2 scroll pumps (Edwards XDS35i). The dilution unit was homemade, but fairly standard: The back-streaming $^3$He gas stream is precooled by the PTC, and then by a cf-hx which uses the enthalpy of the cold gas pumped from the still. The $^3$He is liquefied in the PTC/cf-hx combination (photo of the cf-hx insert in [8]). Then the $^3$He flow is expanded in a flow restriction and cooled in a hx in the still; then it is cooled further by a concentric tube hx and two step exchangers before it is diluted in the mixing chamber.

The flow restriction of the DR which controls the $^3$He flow rate was made from a 0.3 mm i.d. capillary where a 0.25 mm dia. brass wire was inserted. At room temperature, the impedance was 0.3 x 10$^{18}$ m$^{-3}$. The still hx was made from a coiled piece of capillary (0.6 mm i.d., 1 m length) which was placed inside of the still. The concentric tube hx is similar to the one described in [1], but the dimensions have been upped to allow for higher $^3$He circulation rates. To fabricate this hx, a German silver tube (2.5 mm o.d.) was coiled up and placed in a stainless steel tube (5 mm i.d.; 1.5 m length). The step exchangers were made from silver blocks with flow channels and silver sponges for the concentrated and dilute sides. A cross section and a photo of a step hx are depicted in Fig. 3. The mixing chamber consists of a solid silver plate, where 128 silver wires are welded in (photo in Fig. 3). To overcome the thermal boundary (Kapitza) resistance between liquid helium and mixing chamber plate, each silver wire is surrounded by a silver sponge (filling factor ∼ 0.5) with high surface area (10 $m^2$ for each sponge). All silver sponges were made from pressed (500 bar) silver powder [10] which had a surface area of 2.5 $m^2$/g before packing.

More details on this dilution unit can be found in [11]. The refrigeration power of the mixing chamber in this particular configuration was 0.7 mW at 100 mK, and its base temperature was 10 mK.

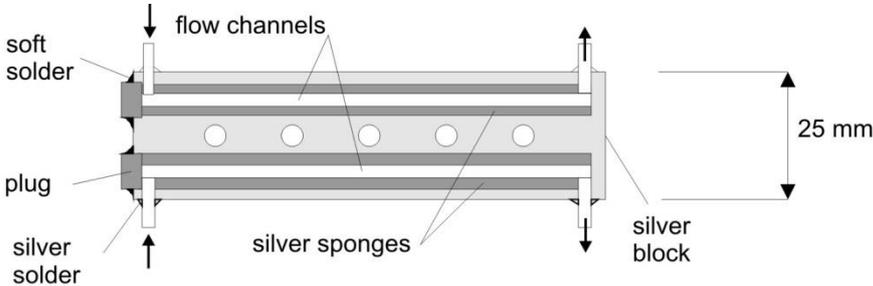

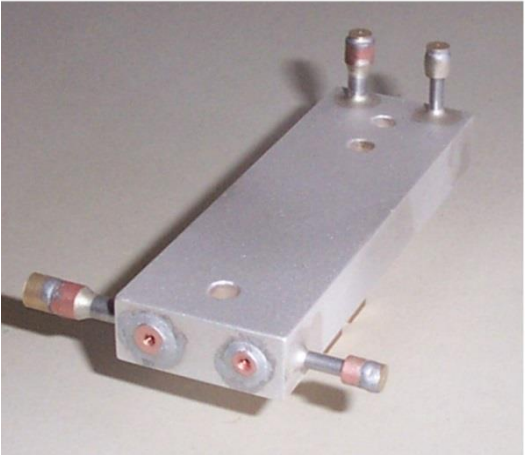
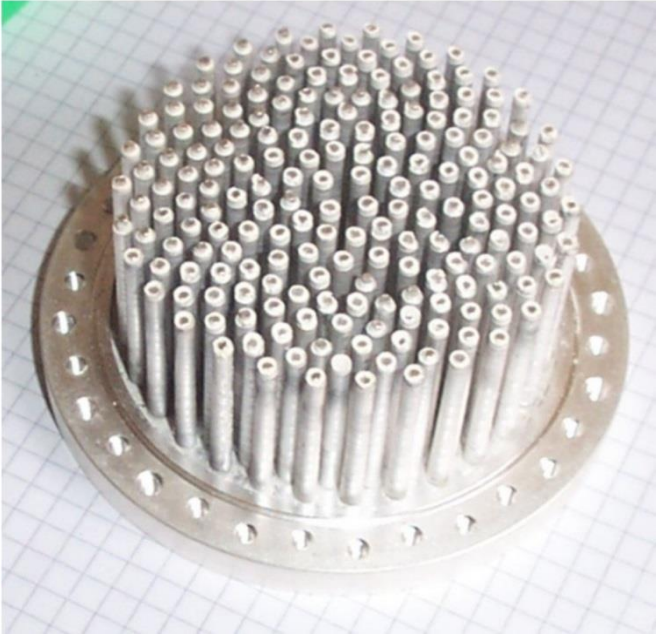

Fig. 3. Top: Cross section of a silver step hx (typical diameter of silver sponge: 5mm). Left side: Step hx, photo (not identical with the hx above). Right side: Mixing chamber bottom plate with silver sponges.

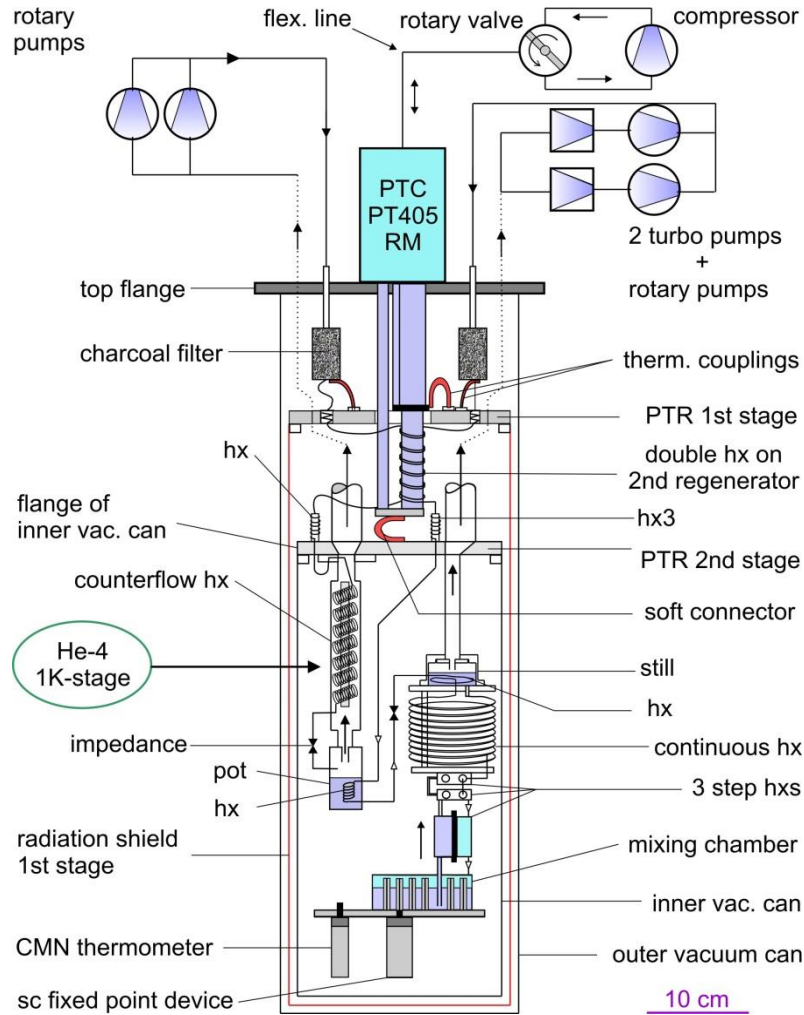

Fig. 4. Cross section of our DR. The 1K-loop is equipped with a cf-hx, and the $^3$He flow of the DR is condensed in a hx in the 1K-pot.

### 4a. Configuration 2: 1K-loop and DR combined

In Fig. 4, the latest version of our DR is shown. The 1K-loop has a cf-hx which, compared to the construction without cf-hx, results in a longer construction. A partial view of the insert of this cf-hx is shown in Fig. 5. To make this insert, a CuNi capillary (0.5 mm i.d.) was coiled up (7 mm i.d.) and then this coil was wound around and soldered to a thin-walled German silver stem; the insert consisted of 18 such windings.

The cooling power of the 1K-stage with cf-hx is given in Fig. 5, and for comparison the cooling power of the 1K-stage without cf-hx (see Fig. 1) is included in the graph.

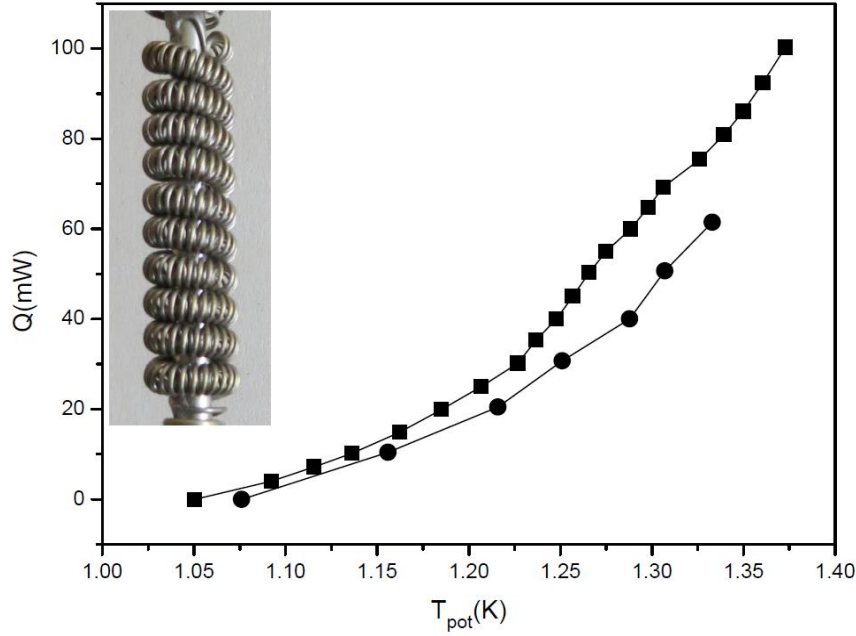

Fig. 5. Refrigeration power of the 1K-stage, with (upper curve) and without cf-hx (lower curve). For both experiments two rotary pumps (Alcatel 2033H) were run in parallel (combined pumping speed: 66 m$^3$/h). Insert: Cf-hx of the 1K-stage of Fig. 4.

A striking change to the 1K-vessel is the addition of a hx where the $^3$He of the DR is condensed and cooled to the vessel temperature of $T_{vess} \sim 1K$. This hx is made from a capillary (0.7 mm i.d.) which is wound around a stud; this stud is inside of the pot and part of the vessel bottom. This modification is especially advantageous during the initial condensation of the $^3$He/$^4$He gas mixture. In our DR, the condensation rate is now 120 std.l/h compared to 60 std.l/h for the DR without 1K-stage. Once the $^3$He/$^4$He is condensed, the 1K-loop can be shut off by pumping its $^4$He back into a storage tank; then $T_{vess}$ approaches $T_{PT2}$ and the DR is run like a regular cryogen-free DR.

However, usually the 1K-loop is not shut off and 1K-loop and DR are run simultaneously. Then the $^3$He flow produces a heat leak $Q_L$ in the vessel of the 1K-loop which is given by the enthalpy difference between $T_{PT2}$ and $T_{vess}$.

$$Q_L / n_3 = H_3(p_{in3}, T_{PT2}) - H_3(p_{in3}, T_{vess}) . \quad (2)$$

In Eq. (2), $n_3$ is the $^3$He flow rate, $H_3$ the enthalpy of $^3$He and $p_{in3}$ the $^3$He inlet pressure.

This heat load $Q_L$ depends heavily on $p_{in3}$ and on $T_{PT2}$. If $p_{in3}$ is below the saturated vapor pressure $p_{3sat}(T_{PT2})$ of $^3$He at $T_{PT2}$, the condensation of the $^3$He flow has to be carried out in the hx of the 1K-pot; the heat load into the 1K-stage will be high. However, if $p_{in3}$ is above $p_{3sat}(T_{PT2})$, the condensation of the $^3$He occurs at the 2$^{nd}$ stage of the PTC (hx3, see Fig. 4) and the heat load into the 1K-pot is much smaller.

This effect is explained in an example; thermodynamic data are taken from [12]: For a typical operating temperature $T_{PT2}$ = 2.8 K, the $^3$He vapor pressure is 0.65*10$^5$ Pa. If $p_{in3}$ = 0.6*10$^5$ Pa, and $n_3$ = 1 mmol/s, we find $Q_L$ = 60 mW. For $p_{in3}$ = 0.7*10$^5$ Pa, we find $Q_L$ = 20.8 mW. This effect can easily be observed experimentally and is shown later on in this report (Fig. 7).

### 4b. Refrigeration power of the still

The cooling power $Q_s$ of the still is given by the enthalpy difference of the $^3$He gas streams entering and leaving the DR, provided contributions of the small fraction of $^4$He in the circulating gas are ignored. Thus, details of the distilling process do not matter in this approximation [13]. $Q_s$ is affected by the installation of a 1K-cooler in the dilution loop. In Fig. 4, where the $^3$He is condensed and cooled to $T_{vess}$ ~ 1K when it enters the dilution unit, $Q_s$ is given by

$$Q_s(T_s) / n_3 = H_{3sat}(T_s) - H_3(p_{in3}, T_{vess}) \qquad (3)$$

Here, $H_{3sat}(T_s)$ is the enthalpy of the saturated $^3$He gas at the still temperature $T_s$. $H_3(p_{in3}, T_{vess})$ in Eq. (3) is small because the $^3$He is in the liquid phase; this is true for all practicable values of $p_{in3}$ and $T_{vess}$. E. g., if $n_3$ = 1 mmol/s, $T_s$ = 0.7 K, $p_{in3}$ = 0.7*10$^5$ Pa, $T_{vess}$ = 1.2 K, then we find a cooling capacity of the still of $Q_s$ = 27.7 mW [12].

For comparison, in Fig. 1 the $^3$He is not condensed in a 1K-stage, but cooled by the 2$^{nd}$ stage of the PTC and a cf-hx, instead. Here, the value of $T_{PT2}$ is critical: If $p_{in3}$ exceeds the vapor pressure $p_{3sat}(T_{PT2})$ corresponding with $T_{PT2}$, condensation takes place in the hx at the 2$^{nd}$ stage of the PTC, and the flow of liquid $^3$He is cooled to $T_s$ in the cf-hx (provided the cf-hx works perfectly). The cooling power of the still is then given by

$$Q_s(T_s) / n_3 = H_{3sat}(T_s) - H_3(p_{in3}, T_s) \qquad (4)$$

The cooling power $Q_s$ in Eq. (4) is thus similar ($Q_s$ = 29.7 mW) to the case above (Eq. (3)) where the $^3$He was condensed in a 1K-stage.

For $p_{in} < p_{3sat}(T_{PT2})$, the $^3$He is not liquefied at the 2$^{nd}$ stage of the PTC. In that case, the cooling power of the still can be found by calculating the enthalpy difference at the hot end of the cf-hx:

$$Q_s(T_s)/ n_3 = H(p_{3sat}(T_s),T_{PT2}) - H_3(p_{in3}, T_{PT2}) \qquad (5)$$

where $p_{3sat}(T_s)$ is the saturated vapor pressure of $^3$He corresponding to the still temperature $T_s$.

If we use the input data of the example given before ($n_3$ = 1 mmol/s, $T_{PT2}$ = 2.8 K, $T_s$ = 0.7K, $p_{in3}$ = 0.6*10$^5$ Pa), we find a still cooling capacity $Q_s$ = 17.8 mW [12].

## 4c. Dilution refrigeration unit

The dilution unit (Fig. 4) was modified in comparison with earlier work (Fig.1) [14]. Most notably, the cf-hx has been removed from the still pumping line (Fig.1), and thus the total length of the dilution unit became shorter. Another step hx was added to the dilution unit to reach lower base temperatures. A commercial CMN thermometer was put in to measure the lowest temperatures of the mixing chamber plate more reliably [15], and a superconducting fixed point device was installed to calibrate the CMN thermometer. (The fixed point device was an older thermometer made by the former NSF (National Science Foundation). Today, a similar device is commercially available [16].)

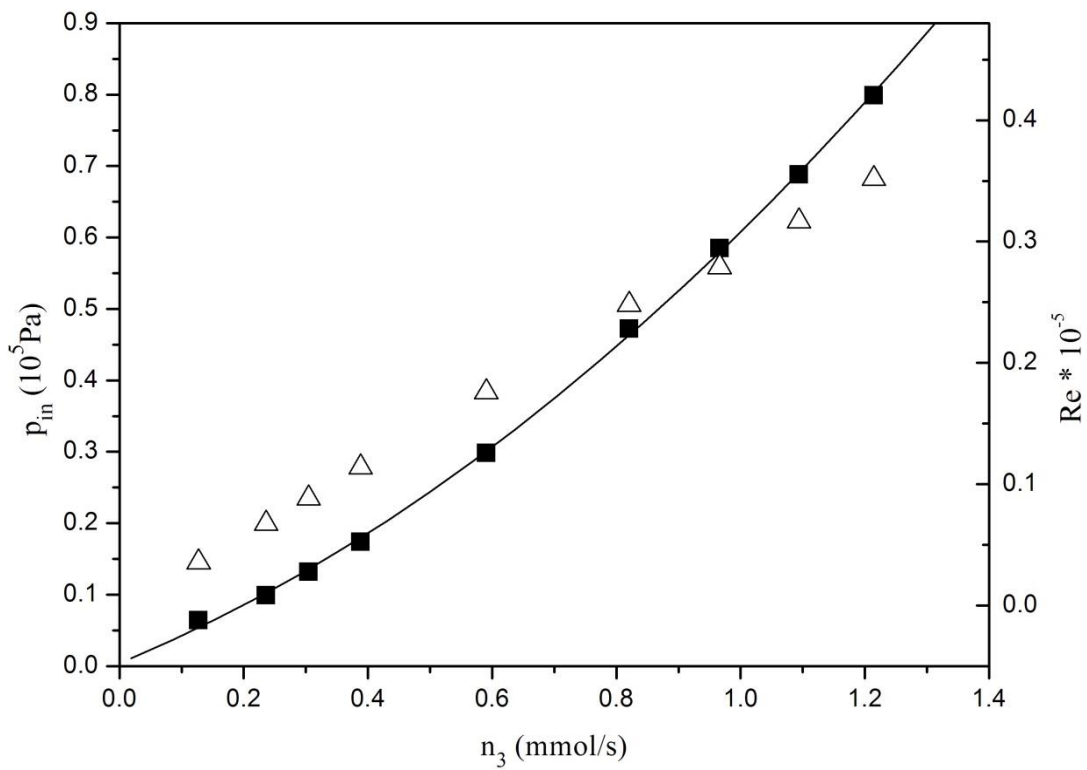

Fig. 6. $^3$He inlet pressure of the DR as a function of the $^3$He throughput (squares, left scale). A quadratic fit has been laid through the data indicating turbulent flow in the flow restriction. Triangles, right scale: Reynolds number, see text.

In Fig. 6, the inlet pressure of the DR is shown as a function of the $^3$He flow rate. With the flow restriction described above we could keep the inlet pressure of the DR below 0.1 MPa for all possible throughputs of the dilution unit and its pumping system. The $^3$He inlet pressure shows a quadratic dependence on the flow rate which could mean that the flow in the impedance is turbulent. For comparison, the Reynolds number Re was calculated [17].

$$Re = G * d_h / \eta \qquad (6)$$

where G is the mass flow rate per unit area, $d_h$ is a hydrodynamic diameter and $\eta$ is the viscosity. G and $d_h$ are known, viscosity data were again taken from [12]. Values for Re are well above a frequently cited critical value of 2300 (Fig. 6) and also indicate turbulent flow.

In Fig. 7, several critical temperatures of the DRs are depicted, namely the temperatures of the 1K-vessel, of the still and of the mixing chamber. The temperature of the 1K-vessel was measured with a RuOx sensor. This temperature curve has a step in its course (marked E in Fig. 7). Starting from a small $^3$He flow rate $n_3$ in Fig. 7, the vessel temperature increased to $n_3 > 0.8$ mmol/s where the corresponding input pressure was $p_{in3} \sim 0.55*10^5$ Pa. In this regime, the condensation of the $^3$He flow occurred in the hx of the 1K-pot. At higher pressures, the 2$^{nd}$ stage of the PTC took over the condensation of the $^3$He flow, and therefore the heat load into the 1K-pot dropped, resulting in a lower pot temperature (see the example at the end of chapter 4a).

The still temperature (also measured with a RuOx thermometer) rose from 0.5 K to 0.8 K with rising $^3$He flow. This is the typical temperature range where the $^3$He distillation is very efficient in the still of a DR [18].

The mixing chamber temperature rose from its lowest temperature of 5 mK at small $^3$He flow rates to 13 mK at the highest circulation rate. The mixing chamber temperature was measured with the CMN-thermometer mentioned above. A graph of the calibration of the CMN magnetization versus the 5 superconducting transition temperatures of our NBS fixed point device is given in [14].

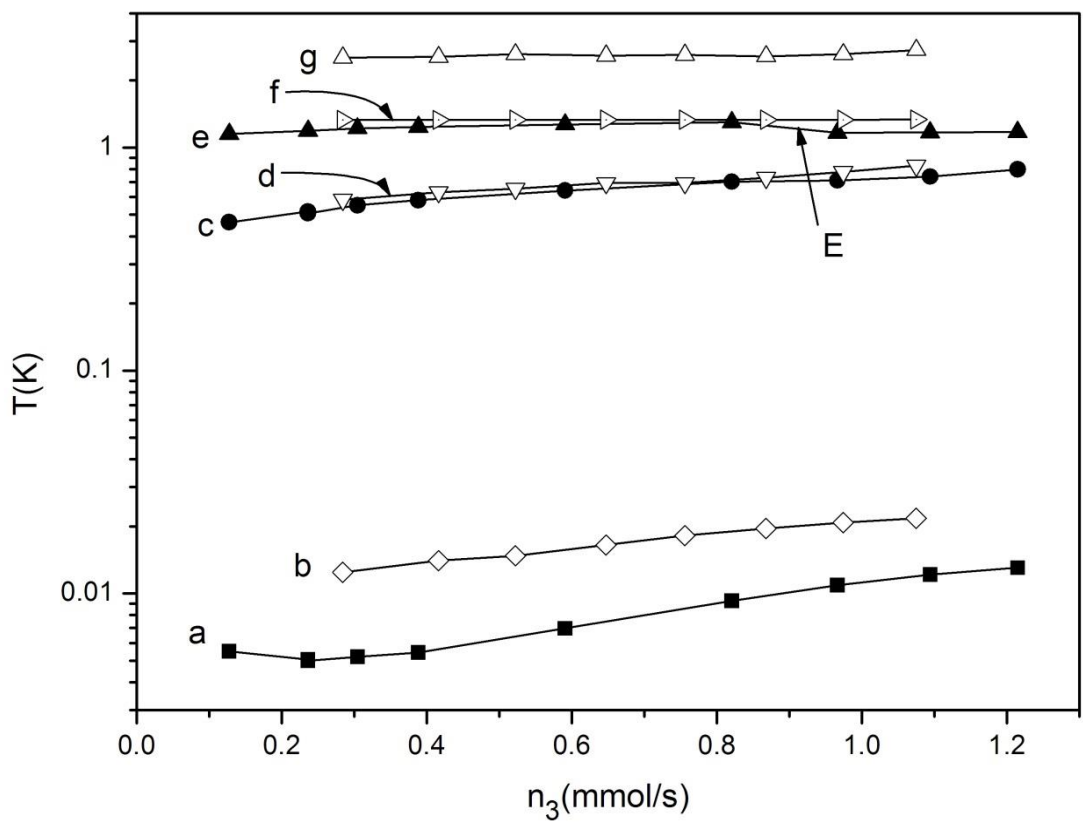

Fig. 7. Various temperatures of our DRs (Figs. 1 and 4) as a function of the $^3$He throughput. a – mixing chamber (Fig. 4); b – mixing chamber (Fig.1); c – still temperature (Fig.4); d – still temperature (Fig.1); e – 1K-stage (Fig.4); f – 1K-stage (Fig. 1); g – $2^{nd}$ stage of PTC (Fig.1). The temperature drop marked E in curve e is reproducible and explained in the text.

## 5. Top loading dry DR with 1K-stage

A very different approach to a DR with 1K-stage has been reported by Kirichek et al. [3,19]. For neutron scattering applications, they combined a 50 mm top-loading cryogen-free cryostat with a $^4$He loop where a base temperature of 1.35 K is reached. Samples can be lowered into the cryostat with a sample stick where exchange gas provides the thermal contact to the cryostat insert. Instead of a sample holder, a commercial dilution refrigeration insert can be installed.

The cryostat is cooled by a Sumitomo RP082 B [20] PTC which has a cooling capacity of 1 W at 4.2 K. The $^4$He inlet line of the 1K-loop is precooled by the two stages of the PTC, and liquid $^4$He is accumulated in a chamber attached to the second stage of the PTC. From there, liquid helium is run through a needle valve to a hx which is in thermal contact with the insert (VTI, variable temperature insert) of the top loading cryostat. The $^4$He which is evaporated in this VTI-hx is pumped by an Edwards XDS35i scroll pump. The authors quote a cooling power of 230 mW at a temperature of the VTI-hx of 1.9 K. The $^4$He vapor pressure at 1.9 K is 2.3 kPa (23 mbar), so in this state the scroll pump is operated at a relatively high inlet pressure. The $^4$He of the 1K-stage is run from a gas cylinder through a nitrogen cooled trap to the cryostat; after having passed the 1K-stage, the helium is pumped into a recovery system. So this is not a closed cycle cryocooler, so far, but it will be converted into one soon. The $^4$He flow rate can be regulated with the needle valve mentioned before. The DR insert is a standard insert from Oxford Instruments (Kelvinox VT) with a base temperature near 25 mK [21].

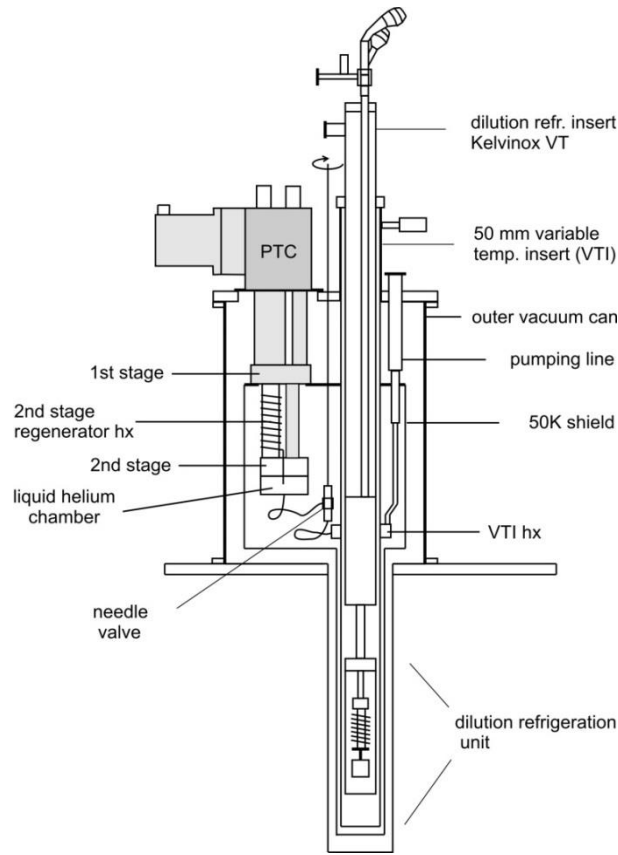

Fig. 8. Top loading cryogen-free DR (courtesy of O. Kirichek [3]).

**Summary and Outlook**

We have described cryogen-free DRs with separate $^4$He-loops whose base temperatures were in the vicinity of 1K; depending on the temperature of their collecting vessel, the refrigeration power was of the order of 100 mW. Two different versions of our cryostat and a top loading cryogen-free 1K cryostat with a commercial DR insert are described. In the latest version of our DR, the dilution loop was run through a hx in the vessel of the 1K-loop so that the condensation of the helium gas stream of the DR was taken over by the 1K-stage. The time to condense the $^3$He/$^4$He mash prior to an experiment was cut in half with the use of the 1K-loop. After condensation, lots of cooling power was available at the vessel of the 1K-stage. In the next version of a dry DR with 1K-stage we plan to add a large mounting plate to the 1K-pot so there is ample room to heat sink amplifiers and electric cables.

Generally, cryogen-free DRs are a mature product. There are several international companies which offer DRs in all sizes necessary for research at competitive prices, standard and customized models. Is there room for further improvements?

There is. We remind the reader of the experimental work on $^4$He circulating DRs [22] where a little fountain pump controls the circulation of the cooling fluid whereas in standard DRs turbo pumps or roots pumps with heavy pumping lines and fore pumps are needed. These pumping systems can get quite big and expensive for powerful DRs and set a limit to cooling powers available at mK-temperatures. Besides, in $^4$He circulating DRs complicated low temperature hxs are not necessary. Instead, the heat exchange at mK-temperatures between concentrated phase and dilute phase is accomplished in a tube where the two liquids flow in opposite directions. They are in direct thermal contact and thus a thermal boundary (Kapitza-) resistance does not occur. But it seems the dynamics of this counterflow hx are not well understood, so far, and more theoretical and experimental work is needed. Lowest temperatures of the mixing chamber of 3.4 mK are promising and encouraging.

Another remarkable attempt to improve dilution refrigeration technology was to build a "completely self-contained cryogen-free DR" where the $^3$He is pumped from the still and re-liquefied at the surface of a cold $^3$He condenser, from where the liquid $^3$He is fed back into the inlet of the dilution unit [23]. The $^3$He condenser is cooled by two $^3$He pots which are pumped by charcoal pumps and alternately either pump $^3$He from a pot or are regenerated. The charcoal pumps are connected to the second stage of a PTC by gas heat switches. This condensation gadget replaces the pumping system that usually comes with standard DRs. Several DRs of that promising new type have been built; a "pocket" version of this type of DR has been introduced in [24]. It seems more effort has to be made in the future to improve on the operating reliability of this type of DR.

An important development is the construction of dry DRs where the closed cycle cryocooler is separated from the dilution refrigeration unit; the thermal connection is made by two perfectly isolated gas lines with circulating helium gas streams [25, 26]. In [25], these gas lines were 2 m long. The motivation for the construction of this type of DR was to keep the vibrations of the cryocooler away from the dilution unit. The spatial separation of cryocooler and experimental setup may also be needed when experiments have to be carried out in screen rooms; the cryocooler would be kept outside of the screen room, and only the DR would be inside. A similar setup has been developed at Cryomech, Inc.[5]. There, the length of the cold gas line is 4 m, but a DR has not been installed at the far end of the gas line, yet.

Quite an interesting new development has popped up just recently which concerned the PTC part of the cryostat. A new type of low frequency linear compressor for the PTC has been reported where a piston moved the helium gas directly into the cryocooler [27], replacing the usually used scroll compressor with rotary valve and motor. The efficiency of the new setup would be greatly improved as losses in rotary valves are on the order of 50 %.

These examples may show that the development of DRs is by no means finished; newly designed cryostats and custom products can be expected in the future.

**Acknowledgements**

The author thanks Oxford Instruments for support and B.S. Chandrasekhar and A. Marx for their interest in this work. K. Neumaier made available several calibrated RuOx thermometers for the dilution refrigeration unit.